# Characterization of particle removal in an airlift pump with a U-bend

Bjørg Synnøve Nigardsøy[1], Vegard Nilsen[1,*], Tom A. Karlsen[2] and Endre Joachim Lerheim Mossige[1]

[1]Faculty of Science and Technology, Norwegian University of Life Sciences, Drøbakveien 31, 1430 Ås, Norway
[2]COWI, Karvesvingen 2, 0579 Oslo, Norway
[*]Corresponding author: vegard.nilsen@nmbu.no

**Abstract**

This paper investigates the self-cleansing performance of an airlift pump with a u-bend. For this purpose, an experimental test model is used to assess the effect of air supply on the pump's ability to lift water and remove particles under different submergence ratios and particle concentrations. In addition, a simple yet accurate fluid mechanical model is used to rationalize the experiments with individual particles, and to predict the critical water velocity required for removal from the pipe bend.

Our experimental results show that the airlift pump is self-cleansing for particle concentrations corresponding to as much as 70% of the cross-sectional area in the u-bend. Furthermore, the self-cleansing ability is relatively independent of the submergence ratio and almost entirely determined by the shear stress. However, the submergence ratio strongly affects energy use, and we find that submergence ratios around 0.75 provide a good compromise between energy use and particle removal efficiency.

**Keywords:** Airlift pump, multiphase flow, fluid mechanics, particle removal, self-cleansing, sewer pipes

# 1. Introduction

Self-cleansing is critically important in sewer pipes to prevent particles from forming deposits that clog up the pipe (Butler et al., 2018), as this can have detrimental consequences such as the accumulation of wastewater in basements. To minimize the risk of clogging, the shear stress acting on the pipe wall needs to exceed a critical value (Ødegaard, 2012); however, this can be hard to accomplish when stormwater and wastewater streams have been separated into different pipes as this reduces the flow rate. Similar problems arise in areas with flat terrain as a result of low hydraulic gradients.

An effective way to increase water flow is to dig deep ditches to increase the hydraulic gradient, but in areas with poor ground conditions such as maritime clay, this requires advanced support structures for retaining loose material during excavation. As this both complicates and increases the cost of the work, it is desirable to avoid digging too deep.

An alternative approach to digging deep ditches for the entire pipeline is to use pumping stations to raise the wastewater pipes along the way (Jones et al., 2006). However, conventional pumping stations are expensive to establish and are associated with high annual costs associated with electricity consumption. This is therefore a rather inefficient use of a pumping station, and especially in cases characterized by relatively small lifting heights. These limitations call for alternative approaches.

Another alternative is to use a so-called airlift pump to pump the water, and with its simple design without any moving parts, it is both robust and reliable, with applications in e.g., sewage treatment plants (Catrawedarma et al., 2021) and aquaculture (Bukhari, 2018), and as a space-saving alternative to deep well pumps (Kassab et al., 2009). Figure 1a shows a possible design for a level lifter in the form of the airlift pump, which is placed in a prefabricated concrete sump or fiberglass-reinforced tank (Karlsen, 2020). The U-bend of the pump enables a direct connection to a wastewater line, and a safe placement of the compressor can be achieved by placing it on an intermediate deck above the pump.
The airlift pump consists of a pipe with an air supply, where the buoyant air bubbles lift the water and the suspended particles up and out of the pump. As such, the working principle of the airlift pump is similar to a commercial coffee brewer, and Figure 1b shows the operation of an airlift pump with a vertical pipe. Due to a suction force resulting from the air injection, water and particles from below the injection point are pulled upwards to replace the amount that is pumped out.

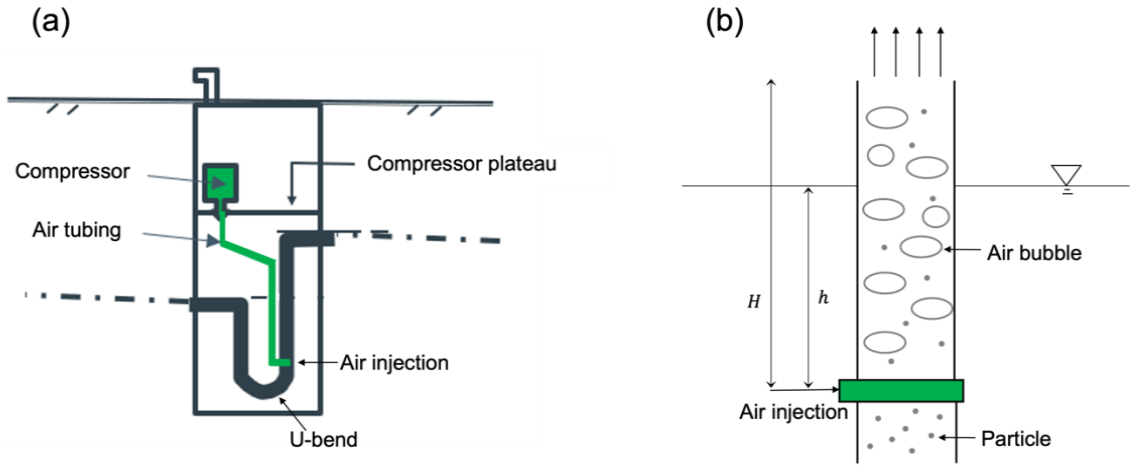

*Figure 1. (a) Possible design of an airlift pump in prefabricated concrete sump or fiberglass-reinforced tank (Karlsen, 2020). (b) Schematic representation of an airlift pump consisting of a single pipe with an air supply in the lower part. The submergence ratio, S_r, is given by the ratio between the height from the air supply and up to the water level, h, and the total height that must be overcome to lift the water with particles out of the pump, H, so S_r = h / H. Figure b is inspired by Catrawedarma et al. (2021)*

In addition to being an energy-efficient (Bukhari, 2018) alternative to pumping stations and other conventional pumping solutions at low lifting heights, the airlift pump has a good ability to transport particles in suspension. Several studies have been conducted to map how particle concentration and particle properties affect the performance of the pump. By combining numerical and experimental data, for example, Mahrous (2012) found that particle removal (measured in kg/h) increased with particle concentration as long as the riser was not clogged. This study also found that the pump transported small particles more efficiently than large ones, which is due to the former having a better ability to follow the liquid flow (due to their lower drag force). Another important study was conducted by Kassab et al. (2007), who through experiments also found that it was more efficient to pump smaller particles than larger ones. In addition, they investigated how different submergence ratios (given as the lift height, $h$, scaled by the total pipe height, $H$, see Figure 1b) affected the particle transport, and it was found that a higher value of this parameter enhanced the particle removal (Tighzert et al., 2013).

An open question, however, relates to the pump's ability to remove sedimentary particles in contact with a pipe wall. To address this important literature gap and to assess its self-cleansing ability, this work utilizes a scaled-down test model with a U-bend. In particular, the experimental results are used to characterize the effects of the air supply on the pump's ability to lift water and remove particles under different submergence ratios and particle amounts. In addition, a theoretical model that rationalizes the single-particle experiments and predicts the flow rate required to remove individual particles is presented. We start with the experimental investigation.

## 2. Methods

The setup consists of a large inlet tank followed by a transparent and flexible u-bend (PVC, diameter: 51mm) that connects to a transparent riser (PVC, diameter: 43mm), see Figure 2a. A hose carries compressed air from the compressor via a regulator and an air flow meter to the air inlet (diameter: 10mm) just above the transparent pipe bend.

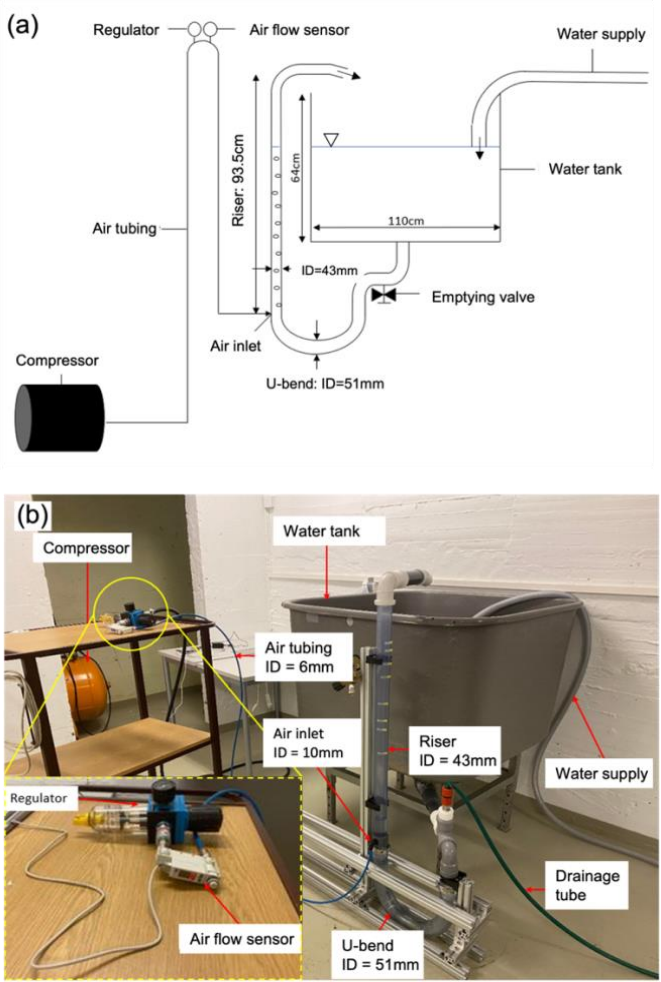

*Figure 2. (a) Schematic representation of the experimental setup and working principle. Air bubbles, provided by the compressor, lift water and particles from the U-bend and riser into the water tank. (b) Picture showing the experimental setup. The submergence ratio is given by the ratio between the water level in the inlet tank, which can be adjusted, and the height of the riser.*

A compressor (EPC 440-250; Eurocomp series) delivers air to the pump via a compressed air hose, see Figure 2b. To regulate the air supply, the air hose is connected to a regulator (MIDI F / R 15 produced; Atlas Copco), which in turn is connected to an air flow meter (PFM711; RS Components) having a limitation of 102 l/min. The regulator and the air flow meter are placed higher than the inlet tank to prevent water from entering the measuring equipment when the compressor is switched off, see inset figure.

The particles used to mimic drainage particles are made of PET (polyethylene terephthalate; Berry Packaging Norway), have a cylindrical shape with a length of 2.8 ± 0.1mm and a diameter of 2.3 ± 0.3mm, and a density of $\rho_p$ = 1395 ± 3.3 kg/m³, see Figure 3.

In order to assess the pump's ability to pump water and remove particles, we conducted independent experiments both with and without particles, as described below.

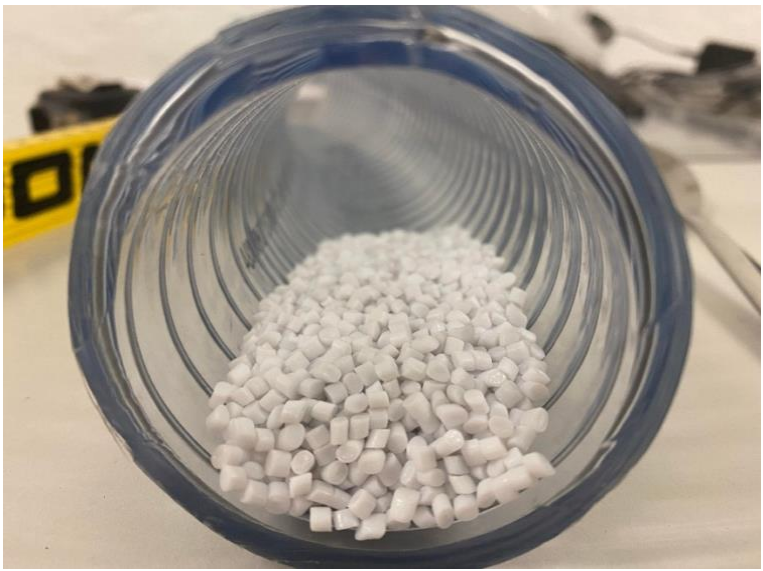

*Figure 3. Particle accumulation consisting of 2032 PET-particles in a flexible PVC pipe. The particles are cylindrical with a length of 2.8 ± 0.1mm and a diameter of 2.3 ± 0.3mm, and have a density of 1395 ± 3.3 kg / m³.*

*Experiments without particles - Calibrating the pump and characterizing flow regimes*
To calibrate the pump and map the effect of air supply and immersion conditions on the water flow, the latter was measured for air rates between 0 and 102 l/min at different S_r values (0.4, 0.5, 0.54, 0.6, 0.65, 0.7, 0.75, 0.8 and 0.82), which are kept constant during each experiment. For each S_r value, the air supply was changed approximately 5-8 times by means of a regulator. At each air supply, at least 3 water flow rate measurements were made

*Experiments with particles - Visualization and characterization of particle transport*

These experiments aimed to identify the critical air supply and associated flow rate of water needed to remove particles over a range of particle amounts (5, 409, 2032 and 11086 particles) and submergence ratios (S_r between 0.4 and 0.82). In a typical experiment, the particles were first introduced into the pipe bend and the air supply was then gradually increased until all particles were removed.

## 3. Results and Discussion

### 3.1 Calibration of the airlift pump without particles

Figure 4 shows how the flow rate of air, $Q_g$, controls the flow rate of water, $Q_l$, under different submergence ratios, S_r. The graphs show that for low values of $Q_g$, $Q_l$ increases rapidly, while for higher $Q_g$ values, $Q_l$ begins to flatten out. This development is due to the fact that the two-phase flow in the riser transitions from a bubbly-slug/slug flow regime (~ 4 l / min - 40 l / min), which is characterized by a high utilization of the air that is supplied to the pump relative to the water flow out of the pump, to a more chaotic flow pattern associated with large viscous losses in the slug-churn and churn flow regime (~ 55 l / min - 70 l / min), as shown in Figure 5. The various flow regimes that can occur in an airlift pump are described by Taitel et al. (1980).

Figure 4 shows that the $Q_g$ required to generate a given $Q_l$ is associated with S_r, which shows that the pump performs best at higher immersion conditions, in accordance with e.g., Kassab et al. (2009). However, as compared with Kassab et al. (2009), our performance curves are shifted to the right, and the water flow, $Q_l$, is somewhat higher. These differences are most likely due to that fact that the pipes used by them were 3.75 m long and with an inner diameter of 25.4 mm, while our pipes are 0.935 m long, with an inner diameter of 43 mm.

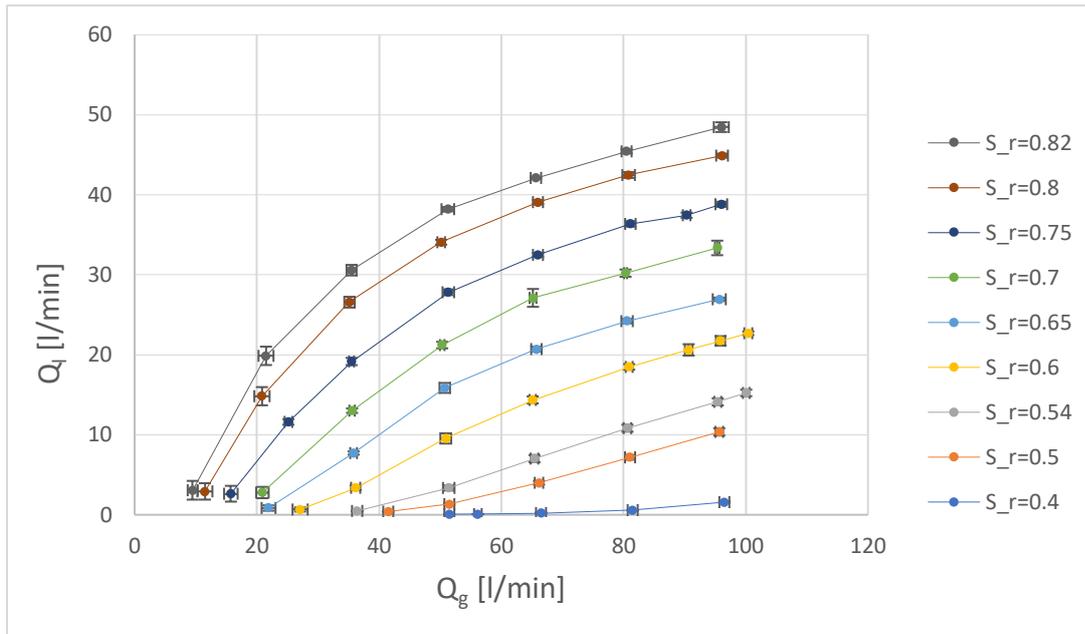

*Figure 4.* Water flow ($Q_l$) plotted against air supply ($Q_g$) at different submergence ratios (S_r). The bars indicate the standard error based on all the measurements.

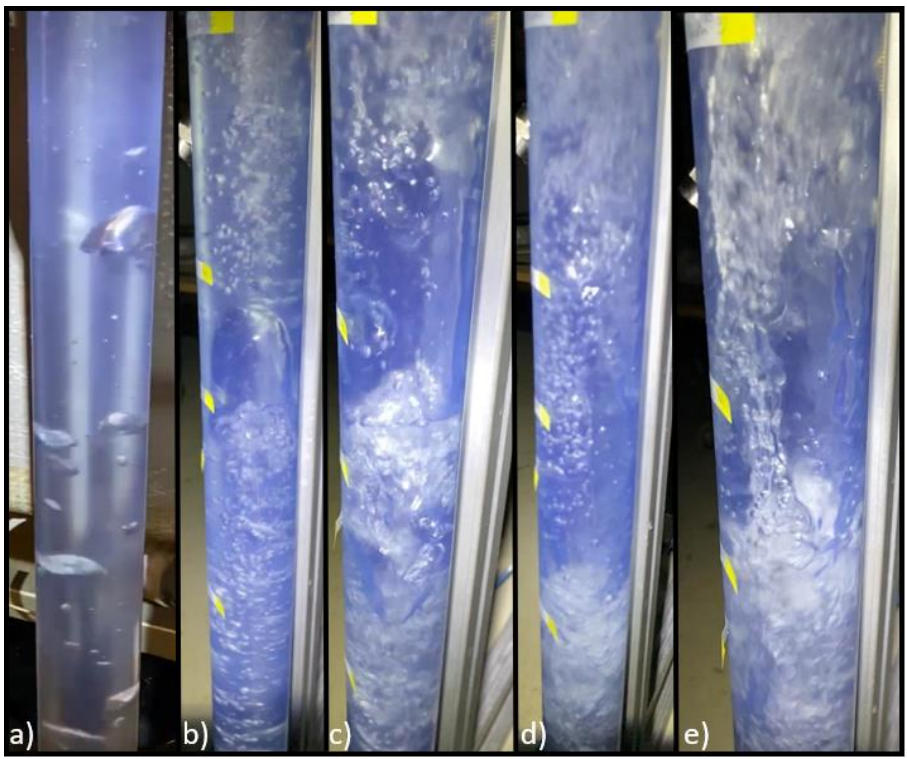

*Figure 5.* Pictures showing different flow regimes in the riser. *a)* bubbly flow ($Q_g$=0-1 l/min), *b)* bubbly-slug flow ($Q_g$=4 l/min), *c)* slug flow ($Q_g$=25 l/min), *d)* slug-churn flow ($Q_g$=55 l/min) and *e)* churn flow ($Q_g$=70 l/min).

For the particle experiments, we are mainly interested in the flow rate of water needed to remove particles from the bend, but it is impractical to measure this quantity and identify particle displacement simultaneously. Therefore, the critical flow rate of water was instead found by reading off the air flow rate from the air flow sensor and using the diagram in Figure 4 to extract the associated water flux. This approach assumes that the particles do not affect the flow rate of water and the overall efficiency of the pump, which is likely to be true for low particle concentrations as used here.

**3.2 Particle visualizations**

The airlift pump consists of a pipe bend and a riser (see Figure 2) and is considered as self-cleansing if particles are removed from both of these pipe parts. However, our experiments show that the critical water velocity is controlled by sedimentary particles in the bend, which is the focus of the visualizations presented below.

By varying the particle concentration between 5 and 11086 particles, three distinct modes for particle displacement in the pipe bend were observed, as illustrated in Figure 6:

    i) Single particle displacement (5 particles)

    ii) "Sand dune displacement" (409, 2032, 11086 particles)

    iii) Plug wise displacement (11086 particles)

In each case, the necessary air flow rate required to remove particles from the pipe was recorded. Particle removal as a function of particle concentration was also recorded through a gradual increase until the critical point was reached.

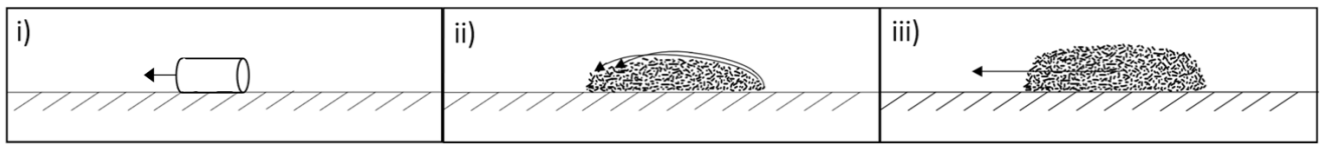

*Figure 6. Schematic representation of the three modes of particle removal from the bend. i) single particles ii) «sand dune displacement» and iii) plug wise displacement. The arrows in i) and ii) indicate the direction of motion of individual particles, while the arrow in iii) shows the displacement of the entire particle aggregate.*

*Mode i):* Individual particles that were not in contact with each other were found to slide along the bottom of the tube wall in response to the flow of water. The critical water velocity generating particle displacement along the bottom was not sufficient to remove them entirely from the pipe, however, as they typically stopped and remained stationary halfway up the u-bend. With a further increase in the flow rate from this point, all the particles were successfully removed.

*Mode ii):* In mode ii) the particles were held together by weak frictional forces, and as a result, higher flow rates were necessary to initiate motion as compared to displacing individual particles in regime i). The displacement of a particle accumulation was reminiscent of "sand dune displacement", see Figure 7b, where the water flow lifts and carries particles to the downstream side of the accumulation, where they are deposited.

A further increase in the air flow rate, which is associated with an increase in water velocity, was necessary to lift the clump of particles higher up in the bend, see Figure 7a. Here, the particles maintained a stable position as the (upward) drag force exerted by the flow of water balanced the (downward) weight of the particles. As the flow rate was again further increased, more and more particles were removed until the number of particles in the particle aggregate was down to zero.

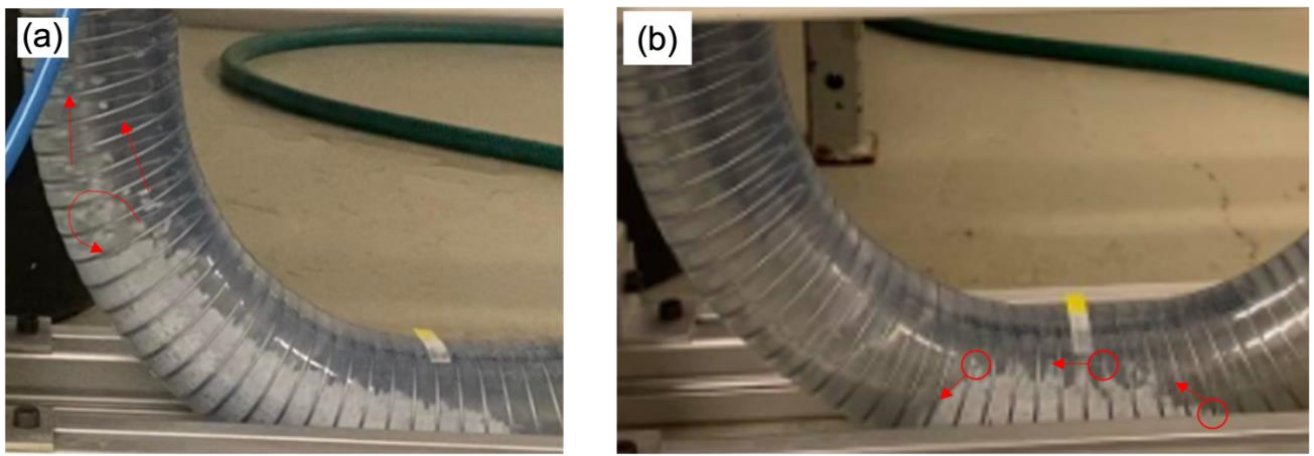

*Figure 7. (a) Displacement of particles in regime ii) in the steepest part of the u-bend (b) Shows detachment and subsequent re-attachment of particles in mode ii) "sand dune displacement".*

*Mode iii):* Finally, the pipe bend was filled with 11086 particles which made up as much as 70% of the cross-sectional area. At start-up, the accumulation of particles was pushed slightly upstream (to the right) as the air was first introduced; this movement was soon followed by a plug wise movement to the left (mode iii)). Thereafter, the particle displacement transitioned into mode ii), but since a larger part of the available flow area was blocked than for 409 and 2032 particles, the local water velocity was higher, which gave a more efficient particle removal of the upper layers. However, this venturi effect diminished as soon as the top particle layers were removed as the available flow cross section increased.

### 3.3 Theoretical model of particle transport in the pipe bend

Several studies have focused on the effect of shape, size and other particle properties on the particle flow performance in an airlift pump, but as far as we know no one has yet developed a theoretical model that describes the motion of particles along the pipe wall in an airlift pump with a u-bend. As such, relevant theory is presented in this chapter to provide a better understanding of the mechanisms underpinning particle removal and to predict the required flow rate of water. To enable a theoretical description, we only consider individual particles, which are primarily advected by frictional drag along the pipe wall, and we ignore secondary transport mechanisms such as particle rotation and lift.

In order to remove particles along the bottom of the pipe bend, the drag force from the water must overcome the friction force between a particle and the pipe wall (Sæterbø et al., 1998), see Figure 8. The friction force depends on the weight of the particle in the fluid, $W$, and the dimensionless coefficient of friction, μ, and is expressed as $F_R = \mu W$.

The particle acceleration $\vec{a}$ is given by Newton's 2. law of motion:

$$m_p \vec{a} = \Sigma \vec{F} = \vec{F}_D - \vec{W} \cdot \mu = \vec{F}_D - \vec{F}_R \qquad (3)$$

, where $m_p$ is the particle mass relative to the surrounding fluid and is expressed as

$$m_p = \frac{W}{g} \qquad (4)$$

If $F_D - F_R > 0$ the particle is accelerated to the left, as shown in Figure 8,

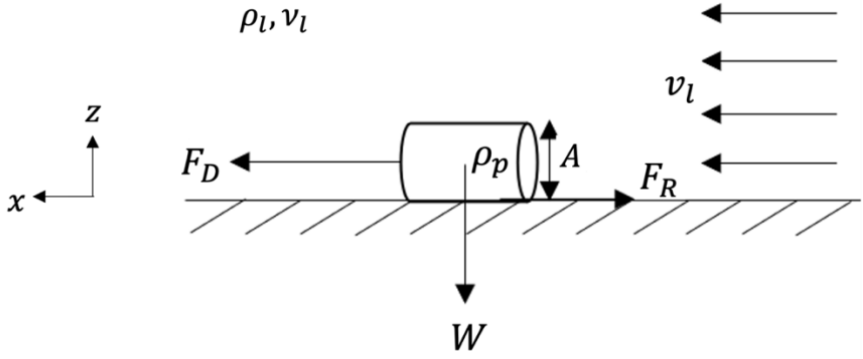

**Figure 8.** Drag ($F_D$), weight (W), and friction force ($F_R$) acting on a single particle along the tube wall in the bend. Since the particle is much smaller than the diameter of the pipe bend and the diameter of the pipe itself, we can safely neglect curvature effects. In order for the particle to be set in motion, the drag force must overcome the friction force against the pipe wall, i.e., $F_D > F_R$. Here, $\rho_p$ is the density of the particle, $\rho_l$ is the density of the fluid (here: water), V is the volume of the particle, and $W = (\rho_p - \rho_l)Vg$ is the normal force of the particle against the substrate. The frictional force against the substrate is given as $F_R = \mu W$, where $\mu$ is the coefficient of friction, with $F_D = c_D A \frac{1}{2}\rho_l v^2$, and v is the water velocity through the tube, and A is the projected area of the particle.

which we can express in terms of the water velocity:

$$v = \sqrt{\frac{2W\mu}{C_D A \rho_l}} \qquad (5)$$

The unknown drag coefficient, $c_D$, is a function of the water velocity, $v$, so Eq. (5) must be solved by iteration, as described in several textbooks (e.g., Elger et al. (2016) and White (1979)). The first step is to guess a value of $v$. Then, $v_l$ is used to calculate the Reynolds number for the flow around the particle, $Re = vd/v_l$, where $d$ is the particle diameter and $v_l$ is the kinematic viscosity of the surrounding fluid ($v_l \sim 10^{-6} m^2/s$ for water). Using the Reynolds number and assuming a disk shape for the cylindrical particles, one then reads $c_D$ from a drag coefficients diagram, see (Lapple and Shepherd, 1940). The drag coefficient is then plugged into formula (5) to find a better approximation of $v_l$, and this iterative process is repeated until the value for $v_l$ converges. In chapter 3.4, we validate the critical water velocities found by this iteration method against experimental data. We also examine the effect of particle shape on the theoretical prediction, and perform theoretical calculations for spherical particles, in addition to the abovementioned disks.

Finally, we calculated the critical shear stress required to remove particles from the pipe wall from the expression below

$$\tau_{wall} = c_f \frac{1}{8} \rho v, \tag{6}$$

see e.g. Elger et al. (2016), where $\rho$ is the density of water and $c_f$ is the friction coefficient found from a Moody diagram (Moody, 1944).

**3.4 Critical water velocity required for particle removal from the pipe bend**

Figure 9 shows the water velocity, $v$, required to remove different amounts of particles from the pipe wall, and the experimental data are compared with the theoretical prediction for individual particles. As we were not able to identify a coefficient of friction between the specific plastic materials of the pipe bend (Polyvinyl chloride) and the particles (polyethylene terephthalate), a general coefficient of friction for plastics against plastics (Tribology), $\mu = 0.4$, was used instead. The results show that by assuming a spherical particle shape we obtain a good approximation, with a difference of only 15% with the experimental data, which is a clear improvement compared to the theoretical prediction using disc-shaped particles.

Figure 9 shows that critical water velocity increases with the number of particles, which is due to the enhanced friction associated with large particle numbers, and because a combination of particle rotation and lift is necessary to initiate transport. However, the results show that critical water velocity flattens out when the particle number exceeds 2000.

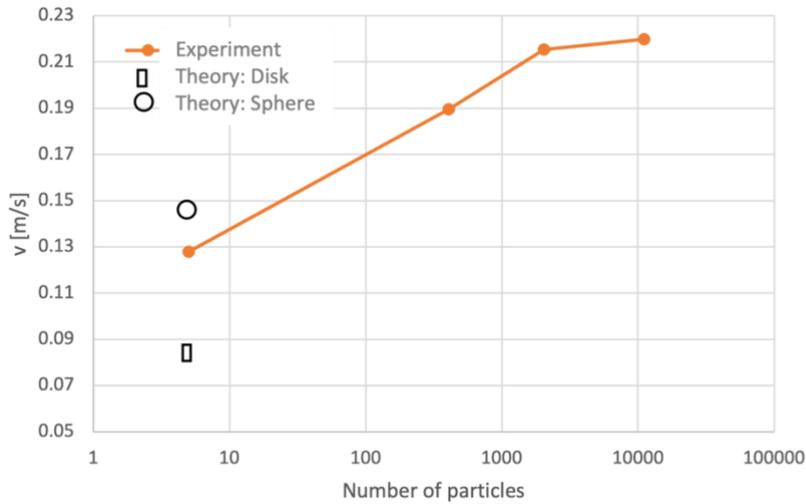

*Figure 9. Critical water velocity required to remove particles from the pipe bend as function of the number of particles. By assuming a spherical particle shape, the theoretical prediction is enhanced as compared to when the particles are approximated as disks. The coefficient of friction was set to µ = 0.4 in the theoretical model. The critical water velocity increases rapidly with the particle number up to 2032 particles, but levels off between 2032 and 11086 particles.*

### 3.5 Shear stress required to remove particles from the pipe bend

As shown in Chapter 3.3, it is possible to predict the motion of individual particles in the pipe bend, but when the particle number increases, complex contact forces, rotation and lift forces can no longer be neglected, which makes this difficult. In such cases, it is much more practical to analyze the granular medium as a whole and to examine the shear force required to initiate particle motion.

Figure 10 shows the mean shear stress at the pipe wall, $\tau_{wall}$, required to remove different amounts of particles from the pipe bend at different submergence ratios, S_r. As expected, the required shear stress increases with the number of particles in the bend, which is largely because the friction between the particles increases with the weight of the granular medium. However, the curves level out in the range of 2032-11086 particles, which is because the water must pass through a smaller cross-sectional area when the particle number increases, which in turn leads to an enhanced local shear stress. It is also worth noting that the immersion ratio is of minimal importance for particle removal as the curves largely coincide.

The required shear stress to remove particles in the bend is in the range 0.07-0.20 N/m², while 2 N/m² is commonly used as requirement for self-cleansing in plastic wastewater pipes (Lindholm (2015)). The large discrepancy indicates that the requirement is too conservative and should potentially be adjusted for airlift pumps, which could reduce operating costs significantly. However, sewage particles tend to form aggregates at solid substrates such as pipe walls (Ashley et al., 1994), so to achieve self-cleansing in real situations, shear stresses beyond 0.20 N/m$^2$ may still be necessary. Also, as the shear stress formula (Eq. 6) was developed for straight pipes it may not predict shear stresses in pipe bends with the same level of accuracy.

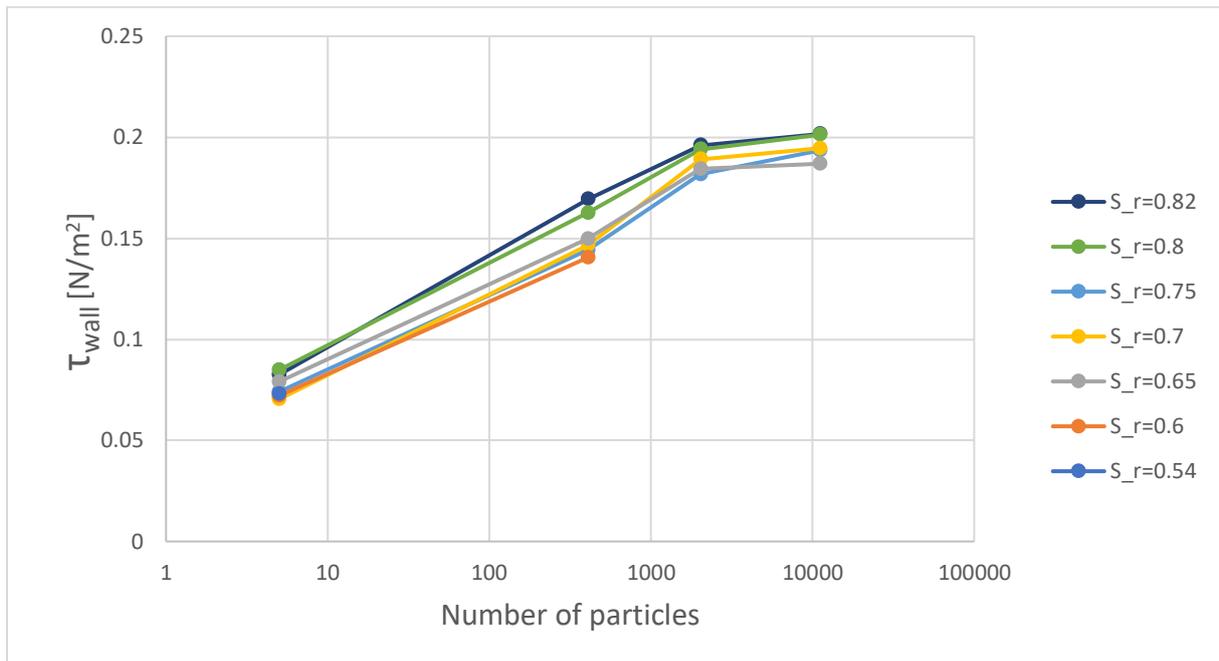

*Figure 10. Wall shear stress required to achieve self- cleansing for different particle numbers and submergence ratios, S_r. The plot shows that the shear stress increases rapidly with the particle number up to 2032, but levels off beyond 2032 particles. The plot also shows that the shear stress is relatively independent on S_r. Standard deviations not shown as they are typically smaller than the symbols used to indicate the mean values.*

### 3.6 Energy use

For pumps, electricity use accounts for a significant proportion of the total operating costs, and it is therefore useful to estimate this parameter in order to compare the airlift pump to other alternatives. Figure 11 shows the energy consumption per time (power) of the air supply, $P_g$, required to remove different particle amounts from the pipe at different immersion conditions, S_r.

As expected, the power required for particle removal increases with the number of particles; this is because the critical water velocity (which depends on the air supply, see Figure 4) and the associated shear force increase with the number of particles, as described in Chapter 3.3 and 3.4, respectively. And as for the critical water velocity and shear force (Figs. 9 and 10), $P_g$ plateaus in the range around 2032-11086 particles.

The submergence ratio, $S_r$, has a strong impact on the energy consumption, which is because more energy is needed to lift a higher column of water. In practice, however, it is not useful to operate the pump under too high submersion ratios as the purpose after all is to raise a water column without digging deep ditches. In addition, Figure 11 shows that the reduction in energy use increases only slightly with $S_r$ beyond $S_r=0.75$. As such, $S_r$ values between 0.7 and 0.75 would potentially make a good compromise between energy costs and construction costs in full-scale implementations.

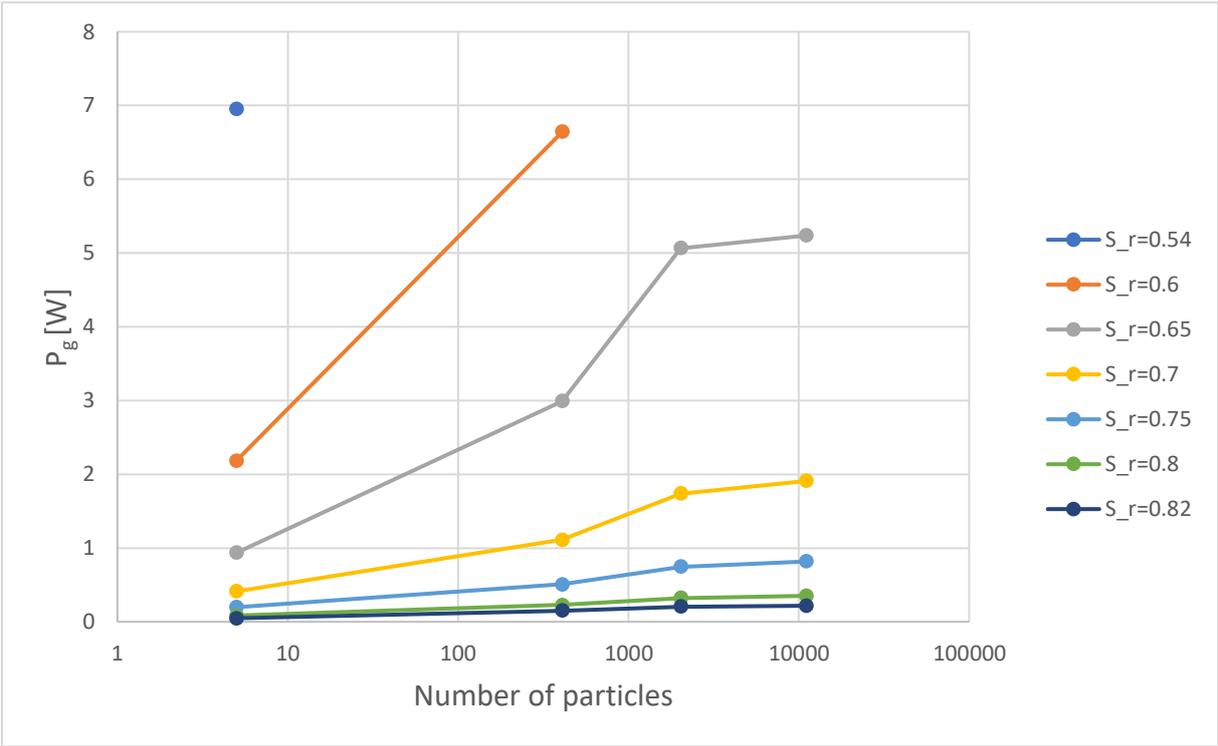

*Figure 11. Energy consumption measured in Watts [W] required to achieve self-cleansing in the pipe bend as function of particle number and submersion ratio, S_r. The energy consumption increases with the particle number and is strongly dependent on S_r up to S_r = 0.7. Standard deviations not shown as they are typically smaller than the symbols used to indicate the mean values.*

## 4. Conclusion

This article investigates the pumping efficiency and self-cleansing potential of the airlift pump using a scaled-down test rig. In particular, the experiments served to characterize the effect of air supply on the pump's ability to lift water and remove synthetic PET particles in a U-bend under different submersion ratios (between 0.54 and 0.82) and particle quantities (between 5 and 11089), corresponding to filling ratios up to 70%. Over this wide range of operational conditions, the pump demonstrated good pumping characteristics and showed a satisfactory self-cleansing performance.

Our results show that the energy consumption was greatly reduced with increasing submersion ratios up to $S_r = 0.75$, and that higher values of $S_r$, which can only be achieved by digging deep ditches, gave only a minimal reduction. Furthermore, the self-cleansing ability of the pump was found to be controlled almost entirely by the shear stress, with the submersion ratio having only a minimal influence.

Finally, a theoretical model was used to predict the water velocity required to remove individual particles from the pipe bend, with a deviation from the experimental data of only 15%. In order to predict the critical velocity required to remove an accumulation of particles from the bend, the model must be extended to account for particle lift and rotation, as well as particle-particle interactions.

## Acknowledgements

We thank Sven Andreas Högfeldt and Øyvind Hansen for fabricating parts used in the experimental setup, Marit Kvalvål Pettersen for providing the synthetic beads, and Jon Arne Engan for critical reading of the master's thesis that laid the basis for this manuscript (Nigardsøy, 2022).